\documentclass[final]{svjour3}
\usepackage{graphicx}
\usepackage{subcaption}
\usepackage{rotating}
\usepackage{amssymb}
\usepackage{mathptmx}
\usepackage[numbers]{natbib}
\makeatletter
\journalname{Journal of Low Temperature Physics}

\usepackage[normalem]{ulem}
\usepackage{color}
\newcommand{\repl}[2]{\textcolor{black}{#2}}
\newcommand{\nb}[1]{}
\newcommand{\done}[1]{}

\bibpunct{}{}{,}{s}{}{,}

\begin{document}

\newcommand{\hdblarrow}{H\makebox[0.9ex][l]{$\downdownarrows$}-}
\title{Operational Optimization to Maximize Dynamic Range in EXCLAIM Microwave Kinetic Inductance Detectors}

\titlerunning{Maximizing Dynamic Range in EXCLAIM MKIDs}
\authorrunning{T.M. Oxholm et al.}

\author{Trevor M. Oxholm$^1$ \and Eric R. Switzer$^2$ \and Emily M. Barrentine$^2$ \and Thomas Essinger-Hileman$^2$ \and James P. Hays-Wehle$^2$ \and Philip D. Mauskopf$^3$ \and Omid Noroozian$^2$ \and Maryam Rahmani$^2$ \and Adrian K. Sinclair$^3$ \and Ryan Stephenson$^3$ \and Thomas R. Stevenson$^2$ \and  Peter T. Timbie$^1$ \and Carolyn Volpert$^4$ \and Eric Weeks$^3$}

\institute{
\email{Trevor M. Oxholm \\ oxholm@wisc.edu} \\
\\
$^1$Department of Physics, University of Wisconsin-Madison, Madison, WI 53706, USA\\
\\
$^2$NASA-Goddard Space Flight Center, Greenbelt, MD 20771, USA\\
\\
$^3$Department of Physics and School Of Earth and Space Exploration, Arizona State University, Tempe, AZ 85287, USA\\
\\
$^4$Department of Astronomy, University of Maryland, College Park, MD 20742, USA
}

\maketitle

\begin{abstract}

Microwave Kinetic Inductance Detectors (MKIDs) are highly scalable detectors that have demonstrated nearly background-limited sensitivity in the far-infrared from high-altitude balloon-borne telescopes and space-like laboratory environments. In addition, the detectors have a rich design space with many optimizable parameters, allowing high sensitivity measurements over a wide dynamic range.  For these reasons, MKIDs were chosen for the Experiment for Cryogenic Large-Aperture Intensity Mapping (EXCLAIM), a balloon-borne telescope targeting nearly background-limited performance in a high-altitude atmospheric environment from 420-540~GHz. We describe MKID optimization in the specific context of EXCLAIM and provide general results that apply to broader applications. Extending the established approach of tone frequency tracking, we show that readout power optimization enables significant, further improvement in dynamic range.


\end{abstract}

\section{Introduction}
MKIDs are pair-breaking superconducting microwave resonators capable of highly sensitive detection of radiation, with applications ranging from millimeter waves to X-rays \repl{}{\citep{day2003,zmuidzinas2012,mauskopf2018}}.  \repl{Large format MKID arrays are}{Arrays of numerous MKID detectors can be} multiplexed on a single transmission line \repl{}{\citep{van2016}} \done{I am not aware of any 'large format' array that uses just a single readout line. Suggest rewording this sentence.} and are highly tunable to maximize sensitivity to a wide range of optical loading, making them ideal for use in terrestrial, suborbital, and space-based telescopes \citep{ulbricht2021}. MKIDs operate through the kinetic inductance effect, whereby energetic photons \repl{incident on}{absorbed in} a superconducting thin film break Cooper pairs\done{this one is very picky - but technically it is only the photons that are absorbed in the film break pairs.}, altering the inductance and resistance of the film. A single feedline can contain a comb of superconducting resonant circuits, each \repl{uniquely}{} \repl{modulated in magnitude and phase in response}{responding in amplitude and phase} to an optical power delivered to the MKID. \done{another minor point, but in principle two KIDs could have exactly the same adulation in phase and amplitude to the same source. I believe I understand the point, but I'd suggest rewording for clarity.}

\done{- Suggest to add a few references in the introduction - in particular at the end of the first and second sentences.}

MKID characteristics are tuned both by the geometry and materials, and dynamically as a function of their readout. The resonator geometry, including the active volume and the coupling capacitance, can be chosen to \repl{minimize noise while maximizing responsivity}{maximize sensitivity}\done{would it be more succinct to say maximize detector sensitivity?}.  The readout system permits additional detector optimization during operation by tuning the frequency and readout power transmitted to each detector. These readout optimizations can be performed uniquely for each detector and optical power, significantly increasing the dynamic range. This tunability is especially advantageous in spectroscopy, where the optical power varies widely across individual channels. \repl{}{In contrast, TESs can be designed to operate at higher optical power by increasing the saturation power through increased leg conduction, and compensating Joule power to target operating conditions. Higher conduction increases the intrinsic phonon noise \citep{mauskopf2018}, making it difficult to achieve operation at high optical power without compromising noise at low optical power.} \done{Can the readout of a TES not also be tuned to increase dynamic range, e.g. by choosing a different bias point - TMO: add calculations describing previous figures we decided not to include? Or bring the TES panel back and make it a 4-panel plot? Emphasize that readpower changes intrinsic noise in the detectors in a similar manner to $G$ in TES detectors, beyond readout operation but $G$ is fixed by geometry in TESs - mention in email to Phil.}

EXCLAIM \citep{cataldo2020} is a balloon-borne cryogenic telescope featuring an aluminum MKID array and designed to demonstrate the line intensity mapping technique \citep{Kovetz2017} to obtain tomographic maps of extragalactic carbon monoxide and singly-ionized carbon emission, which may be used to infer the cosmic star formation history \citep{carilli2013}. EXCLAIM features a set of six $\mu$Spec spectrometers-on-a-chip with resolving power $R=512$ read out in a 3.25-3.75 GHz microwave band\citep{mirzaei2020} by a Xilinx RFSoC. \repl{We assume a $3.50\,\rm GHz$ tone throughout this manuscript.}{} \nb{this is now given in the table as a design choice}

MKIDs have already demonstrated nearly background-limited performance in balloon \citep{masi2019,gordon2020} and space-like \citep{baselmans2017} backgrounds in the far-infrared. Like other balloon- and space-based mid- to far-infrared missions\done{suggest add a reference or two}, EXCLAIM detectors must accommodate a wide range of background loads spanning three orders of magnitude. MKIDs proposed for next-generation space instruments\citep{glenn2021,leisawitz2021} require high sensitivity over loads ranging \repl{more than}{from three \citep{glenn2021} to} five \repl{}{\citep{bradford2021}} orders of magnitude, owing to the wide \repl{range}{variety} of science cases targeted by a single instrument\repl{\citep{bradford2021}}{}. \repl{}{While the EXCLAIM mission does not require sensitivity to the brightest background loads, it will provide a valuable testbed for dynamic range optimization for future far-IR missions.}

This study describes the optimization of the EXCLAIM MKID design over a wide range of background loads and the underlying device physics for general MKID applications. We focus on optimizing the readout system \repl{}{with a purely model-based approach}, which can support dynamic range \repl{}{optimization} during operation. \repl{}{Throughout, we simplify the model by choosing a single signal modulation frequency of $1\,\rm Hz$ and constant quasiparticle pair-breaking efficiencies, and we describe how these effects may be accounted for in the discussion section.}

\section{MKID sensitivity to a widely-varying background}
\label{sec:modeling}

\begin{table}[]
    \centering
    \begin{tabular}{|c|c|c|c|}
    \hline
    \textbf{Parameter} & \textbf{Symbol} & \textbf{Parameter type} & \textbf{Value} \\
    \hline
    \hline
    \textbf{Resonator volume} & $V$ & Design choice & 374 $\mu {\rm m}^3$ \\
    \textbf{Dark resonator frequency} & $\nu_0$ & Design choice & 3.5\,GHz \\
    \textbf{Coupling quality factor} & $Q_c$ & Design choice & $2.3 \times 10^5$ \\
    \textbf{Residual quality factor} & $Q_{i0}$ & \repl{Measured}{Material} & $1.75\times 10^6$ \\
    \textbf{Kinetic inductance fraction} & $\alpha$ & \repl{Measured/designed/material}{Material/Design} & 0.775 \\
    \textbf{Readout qp generation efficiency} & $\eta_{\rm read}$ & \repl{Measured}{Material} & $9.24\times 10^{-4}$ \\
    \textbf{Amplifier temperature} & $T_{\rm amp}$ & Design choice & 4.1 K \\
    \textbf{TLS exponent} & $\alpha_{\rm TLS}$ & \repl{Measured}{Material} & -0.69 \\
    \textbf{TLS spectral density at 1 kHz} & $S_{0,\rm TLS}$ & \repl{Measured}{Material} &  $1.49 \times 10^{-16}$ Hz$^{-1}$ \\
    \textbf{TLS photon number} & $N_{\rm TLS}$ & \repl{Measured}{Material} & 241 \\
    \textbf{Critical temperature} & $T_c$ & \repl{Measured}{Material} & $1.33\,\rm K$ \\
    \hline
    \textbf{Bath temperature} & $T_{\rm bath}$ & Operational & 100 mK \\
    \textbf{Read tone frequency} & $\nu_{\rm read}$ & Operational & variable \\
    \textbf{Read tone power} & $P_{\rm read}$ & Operational & variable \\
    \hline
    \end{tabular}
    \caption{MKID design reference parameters for the EXCLAIM detector array. \repl{}{MKID performance is determined by the resonator film's material parameters, design choices, and operational parameters. Material parameters in our model are based on lab measurements of Al CPW resonators fabricated at NASA-Goddard. For simplicity, we assume that the bath temperature is fixed.} \nb{There were two copies of the modulation frequency; I've removed both, as this is given in the text already and is a science parameter not a parameter related to the detector/readout.}}
    \label{tab:params}
\end{table}
\vspace{-2mm}

The anticipated incident background load for EXCLAIM ranges from $P_{\rm opt} {\approx} 0.1$ to $100\,\rm~fW$ at the input of the spectrometer throughout the $420{-}540\,\rm~GHz$ passband, with strong frequency-to-frequency variation driven by narrow atmospheric emission lines subject to low pressure broadening in the upper atmosphere \citep{cataldo2020}. Once the flight stabilizes, observations occur at $45^\circ$ elevation at an altitude of ${\sim} 34\,\rm~km$. Hence, dynamic range requirements apply to temporally stable loading rather than requiring significant real-time response. The space background is one to two orders of magnitude below the dark windows in upper atmospheric emission in the EXCLAIM band. Light passes through a cold stop to \repl{control spill}{minimize stray light}\done{very much a colloquialism, suggest rewording for a wider audience}, then through the lenslet-coupled spectrometer, then into the MKID array. We estimate antenna efficiency (to the input of the spectrometer formed by the cold stop) $\eta_{\rm ant} = 0.85$ and detector efficiency (through the cold stop to the detectors) of $\eta_{\rm det} = 0.23$. Throughout, $P_{\rm opt}$ and NEP are defined at the cold stop (incident on the spectrometer lenslet) rather than at the detector in the on-chip spectrometer.

Assuming a model similar to \citet{zmuidzinas2012} and \citet{mauskopf2018} with MKID design parameters shown in Tab.\,\ref{tab:params}, we calculate the total NEP per detector through
\begin{equation}
    {\rm NEP}_{\rm tot}^2 = {\rm NEP}_{\rm opt}^2 + {\rm NEP}_{\rm gen}^2 + {\rm NEP}_{\rm rec}^2 + {\rm NEP}_{\rm amp}^2 + {\rm NEP}_{\rm TLS}^2,
\end{equation}
representing the total, optical, quasiparticle generation and recombination, amplifier, and two-level system (TLS) noise, respectively. Throughout, we assume frequency readout and signal modulation at a nominal frequency $f_{\rm mod} =1\,\rm Hz$. The dominant sources of noise depend on the level of background radiation, which we specify in three regimes:

\textbf{Low-background Loads}: for low absorbed optical power, the detector NEP is typically dominated by generation noise and TLS noise, where the latter only affects readout in the frequency direction. Generation noise is caused by an increase in the number of quasiparticles from thermal phonons and readout photons, while TLS noise is produced by two-level systems at the boundaries of dielectric layers in the film. The NEP of these contributions is
\begin{equation}
        {\rm NEP}^2_{\rm gen} + {\rm NEP}^2_{\rm TLS} = 4 \left ( \Gamma_{\rm th} + \Gamma_{\rm read} \right ) \left ( \frac{d\Gamma}{dP_{\rm opt}} \right )^{-2} + S_{\rm TLS} \left ( \frac{d x}{d P_{\rm opt}} \right )^{-2},
        \label{eqn:NEP_gen_TLS}
    \end{equation}
where $\Gamma_{\rm th}$ and $\Gamma_{\rm read}$ are the quasiparticle generation rates due to thermal phonons and readout photons, respectively, and $dx$ is the differential fractional frequency shift due to changes in optical power.
We model the readout generation rate as $\Gamma_{\rm read} = \eta_{\rm read} P_{\rm read}^{\rm abs}/\Delta$, with $P_{\rm read}^{\rm abs}$ the absorbed readout power and $\Delta = 1.764 k_B T_c=180\,\rm meV$ the gap energy, where $T_c$ is the critical temperature. We assume a constant $\eta_{\rm read}$, though generally it may depend on the optical and readout powers (see Discussion below). Within the TLS noise term, $S_{\rm TLS}=S_{0,\rm TLS} (f_{\rm mod}/\nu_{\rm TLS})^{\alpha_{\rm TLS}} \delta_{\rm TLS}$ is the TLS power spectral density with $\nu_{\rm TLS} = 1\,{\rm kHz}$ pivot, and $\delta_{\rm TLS} = \tanh [ h\nu_{\rm read}/ (2 k_B T)]^{1.5} \cdot (N_{\rm ph}+N_{\rm TLS})^{-1/2}$ describes losses due to TLS, with $N_{\rm ph} = P_{\rm read}^{\rm abs} Q_i (2 \pi h \nu_{\rm read}^2)^{-1}$ the number of readout photons. Here, $\nu_{\rm read}$ is the microwave readout frequency and $h$ is Planck's constant. This form matches measurements of resonators with $\nu_{\rm read}=3.4\,\rm GHz$ by \citet{gao2008} suggesting $S_{\rm TLS}\sim T^{-1.5}$ for $T>100\,\rm mK$. $d x / d P_{\rm opt} \equiv R_x \propto \tau_{\rm qp}/V$ is the frequency responsivity describing variations in resonator frequency due to the kinetic inductance effect, and $d\Gamma/dP_{\rm opt} = q/h\nu$, where $q\equiv h\nu\eta_{\rm pb}/\Delta$ is the number of quasiparticles produced per photon, and $\eta_{\rm pb}$ describes the pair-breaking efficiency, evaluating to $\eta_{\rm pb}=0.57$ when the photon energy $> 2 \Delta$, as is the case for the optical frequencies of interest here \citep{kozorezov2000}. \done{while this is probably a sufficient assumption for this analysis, it is worth considering that the functional form of $\eta_{\rm pb}(\nu)$ around 4-delta ($\sim400\,\rm GHz$) is not constant. Also, as shown by Guruswamy et al. 2014, the limiting value at high energy is strongly dependent on the specific mechanism that governs energy downconversion/loss. TMO: addressed in Discussion section along with discussion on kinetic equations.}
Note that noise from single-quasiparticle interactions may also contribute to generation noise due to e.g. magnetic field flux trapping \citep{flanigan2016}.

\textbf{Photon Background-limited Loads}: the background-limited noise-equivalent power (NEP) for unpolarized radiation at the input of an incoherent detector such as an MKID is \citep{zmuidzinas2003}
\begin{equation}
        \rm NEP_{opt}^2 = 2 h \nu P_{opt} + \frac{2 P_{opt}^2}{B}, 
        \label{eqn:NEP_bknd}
\end{equation}
where $\nu$ is the frequency of the incident radiation, $B$ is the spectrometer optical bandwidth per channel, and $P_{\rm opt}$ is the incident optical power.  

Recombination noise due to optically-generated quasiparticles produces an additional irreducible source of noise \citep{lowitz2014}. ${\rm NEP}_{\rm rec}^2 \approx 4 h \nu q^{-1} P_{\rm opt}$ in the case that quasiparticle generation is dominated by optical photons. For EXCLAIM we find that ${\rm NEP}_{\rm rec}/{\rm NEP}_{\rm photon} \approx \sqrt{2 q^{-1}}$, producing a ${\sim} 23\%$ increase in NEP compared to the photon background at $480\,\rm GHz$. 
    
\textbf{High-background Loads}: for high optical power, amplifier noise dominates the EXCLAIM detector noise, with NEP given by
    \begin{equation}
        {\rm NEP}_{\rm amp}^2 = \frac{k_B T_{\rm amp}}{P_{\rm read}^{\rm feed}} \left ( \left | \frac{\partial S_{21}}{\partial x} \right | R_x \right )^{-2} = \frac{k_B T_{\rm amp}}{P_{\rm read}^{\rm feed}} \left ( \chi_a Q_i R_x \right )^{-2}.
        \label{eqn:NEP_amp}
    \end{equation}
where $\chi_a = 2 (Q_r^2 Q_i^{-1} Q_c^{-1}) \left ( 1 + 2 Q_r x \right )^{-1}$
is the absorption efficiency, with $x$ the fractional frequency detuning, which equals zero when the readout tone frequency is on resonance.
Here, the resonator quality factor is $Q_r^{-1}=Q_c^{-1}+Q_i^{-1}$, and the $\chi_a$ is maximized when $Q_i=Q_c$. $\chi_a$ represents the fraction of the delivered readout power $P_{\rm read}^{\rm feed}$ that is absorbed by the MKID, as $P_{\rm read}^{\rm abs} = \chi_a P_{\rm read}^{\rm feed}$.

\section{Readout tone-tracking and readpower optimization}

\begin{figure}
    \centering
    \includegraphics[width=\textwidth]{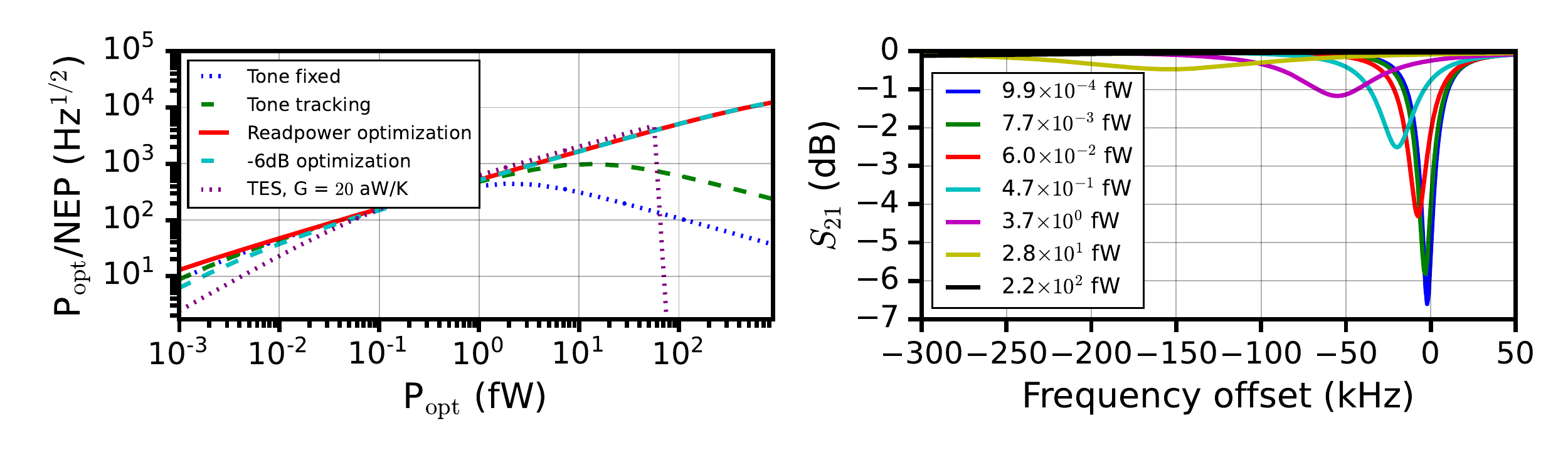}
    \caption{\repl{C}{Simulated c}omparison between the three operational regimes. Left: instantaneous dynamic range, defined as the ratio of the optical power to the total NEP, akin to the signal-to-noise for one second of integration time. Right: magnitude of transmission $S_{21}$ for a selection of incident optical power across the EXCLAIM band.}
    \label{fig:2-panel}
\end{figure}

\begin{figure}
    \centering
    \includegraphics[width=\textwidth]{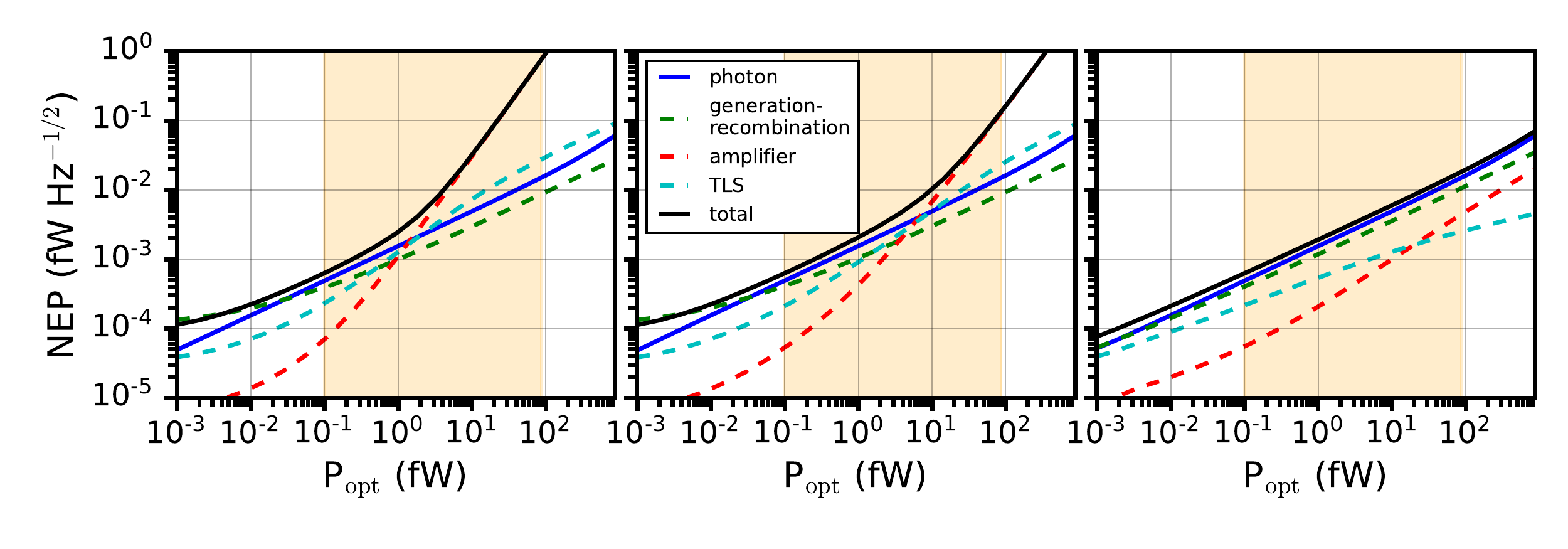}
    \caption{\repl{}{Simulated} NEP curves for three cases: (left) fixed readpower without tone-tracking, (center) fixed readpower with tone-tracking, and (right) optimized readpower with tone-tracking. The shaded orange region represents the range of anticipated incident optical power for the EXCLAIM mission. In the left two plots, the readpower is fixed to minimize NEP at $P_{\rm opt} = 0.15\,\rm fW$, whereas the right plot optimizes at all input optical powers. In the left plot, the lack of tone-tracking causes a frequency mismatch, thereby increasing amplifier noise at high input optical power.}
    \label{fig:NEP_comps}
\end{figure}

Several design parameters (e.g. $V$ and $Q_c$) may be used to optimize detectors for a given application but are functions of the resonator geometry, and therefore cannot be optimized operationally. On the other hand, the readout system may implement in-situ NEP optimization as a function of optical power. The readout system enables two techniques:

\textbf{Resonance tone-tracking}: the tracking of each readout tone to be exactly on-resonance, i.e. $x {=} 0$ for $\chi_a$ in Eqn. \ref{eqn:NEP_amp}, minimizing amplifier noise. Without tone-tracking, we assume the readout tone equals the resonance frequency under a nominal optical loading and that the resonator can move away from this fixed tone under changing optical power (Fig.\,\ref{fig:2-panel}, right).
Note that a similar dependence occurs when reading out in dissipation quadrature, where ${\rm NEP}_{\rm amp,diss} \propto \chi_a^{-1}$.
    
\textbf{Readout power optimization}: an adjustable readout power $P_{\rm read}^{\rm feed}$ delivered to each detector, minimizing the NEP per frequency channel. For low optical powers, a decrease in $P_{\rm read}^{\rm feed}$ leads to a decrease in ${\rm NEP}_{\rm gen}$, provided the resonator is cold enough for thermal quasiparticle generation and TLS noise to be subdominant. For high optical powers, an increase in $P_{\rm read}^{\rm feed}$ suppresses amplifier noise, up to the regime where the resonator response begins to bifurcate \citep{swenson2013} or otherwise display signs of nonlinear response.

These in-situ optimizations are significant for EXCLAIM because they maximize the sensitivity per channel without uniquely fabricating each detector to its anticipated flight loading.
In EXCLAIM's expected flight operation, the tone-tracking and readpower optimization steps will be performed once the altitude stabilizes \repl{}{through measurements of quality factor and resonator frequencies as a function of several readpowers, requiring roughly a minute of data. We anticipate only requiring this optimization once, after the altitude stabilizes. Approaches for real-time optimization of readpower in response to optical power are deferred to future work, but may measure derivatives between tones on and near resonance}. A \repl{}{low-power} pulsed optical reference emitter will enable \repl{ongoing}{periodic} responsivity checks\repl{}{\citep{switzer2021}}. \done{how is this signal coupled in to the detector? Do you foresee any impact on the responsivity in adding this signal, especially in the low background case? Cite JATIS paper - Herschel-SPIRE, BLAST.} \repl{}{Within the proposed RFSoC-based readout design for EXCLAIM \citep{stephenson2022}, a PID loop style tone tracking algorithm can be applied after the digital down conversion stage in a time-division multiplexed fashion.  Two PIDs will be used, each with an estimated resource utilization of 3 DSP48 multipliers, 2 FIFOs of length 512, and 5 adders. This uses only a fraction of the resources compared to the front-end FFT.}

In the low optical power regime where noise is dominated by quasiparticle generation-recombination and TLS, the readpower that minimizes noise also maximizes the internal quality factor $Q_i = \left ( Q_{\rm qp}^{-1} + Q_{\rm TLS}^{-1} + Q_{i0}^{-1} \right )^{-1}$. Here, $Q_{i0}$ is an empirical residual quality factor, $Q_{\rm TLS}=2.61\times 10^{4} \delta_{\rm TLS}^{-1}$ is the dissipation due to two-level systems with $\delta_{\rm TLS}$ defined after Eqn. \ref{eqn:NEP_gen_TLS}, and $Q_{\rm qp}\propto \delta n_{\rm qp}^{-1}$ is the dissipation due to quasiparticles in the film, which decreases with an increased number of quasiparticles. Note that readpower optimization differs from the typical operational procedure to maximize $\chi_a$ by setting $Q_i=Q_c$. Utilizing readpower optimization, a design with $Q_c$ lower than $Q_i$ can achieve near-ideal noise performance while also providing robustness to changes in optical loading, realized quality factors, and resonance-finding.
Because $Q_i$ can be measured directly through $S_{21}$ in a readpower sweep, this technique is less time- and computation-intensive compared to measuring and minimizing the noise directly. The relation between maximal $Q_i$ and minimal NEP breaks down for high optical powers where amplifier noise dominates; in this case, increasing the readpower reduces the noise. For EXCLAIM, however, these optically bright, high-noise channels do not contribute as strongly to the extragalactic science signal, so the $Q_i$ optimization step will suffice. More work will be needed to define the readpower optimization routine in future missions requiring higher sensitivities to brighter sources.
    
Based on these techniques, we analyze the noise performance of the EXCLAIM detector design under three regimes: (i) fixed tone (i.e. the tone frequency is fixed to the resonance under nominal loading) with fixed readpower; (ii) resonance tone-tracking with fixed readpower; (iii) resonance tone-tracking with optimized readpower. While the first two regimes have been demonstrated in laboratory and operational environments \citep{hoh2020}, the active readout power optimization represents a new technique; previous approaches have optimized the readout power by e.g. setting it to $-6\,\rm dB$ below the bifurcation power threshold \citep{gordon2020}\repl{, which can lead to higher noise at lower optical power}{}. \repl{}{In the model presented here, setting the feedline readpower $-6\,\rm dB$ below the bifurcation threshold and including tone tracking leads to a 26\% increase in NEP compared to our optimization method for $0.01\,\rm fW$ loads and a 108\% increase for $10\times$ lower loads, while $1000\,\rm fW$ loads are within 1\%.} 

Fig. \ref{fig:2-panel} (left) shows the dynamic range performance and Fig.\,\ref{fig:NEP_comps} shows the noise contributions in each operational regime\repl{}{, with minimum loads set a decade lower than the EXCLAIM minimum}. In the case of the fixed readpower, $P_{\rm read}^{\rm feed} = 4.8\,\rm fW$ ($P_{\rm read}^{\rm abs} = 2.4\,\rm fW$) minimizes the NEP at $P_{\rm opt} = 0.15\,\rm fW$, representing a typical dark channel in the EXCLAIM passband.
The case with optimized readpower scales approximately as $P_{\rm read}^{\rm feed} \approx 1.6~{\rm fW} + 27~ P_{\rm opt}$. \repl{}{A basic TES sensitivity model is also shown, including noise from phonons and photons, as well as saturation. Here, we follow the model of \citet{mauskopf2018} taking $G=20\,\rm aW/K$ and $F_{\rm link} = 0.6$.}

\section{Discussion}

We have investigated and modeled the physical effects of MKID optimizations through the readout system, including resonance tone-tracking and the optimization of the readout power as a function of optical power.
These conclusions have several caveats that we will study through future measurements. In particular, nonlinear behavior has been observed in MKIDs with high incident background loads, due to bifurcation and nonlinear heating. In Goddard Al CPW test devices, this behavior has tentatively been observed at higher readpowers than the range described here. Future models can employ quasiparticle kinetic equations\citep{chang1978,goldie2012} representing the local heating and cooling of quasiparticles\repl{.}{, resulting in non-constant values for $\eta_{\rm read}$ and, to a lesser extent, $\eta_{\rm pb}$ \citep{guruswamy2014}.} The simple model used here agrees with readpower sweeps in the critical regime across the maximum of the quality factor.

Second, the quasiparticle lifetime may limit sensitivities at lower power levels and rapid signal variations. As a function of signal modulation frequency $f_{\rm mod}$, the finite quasiparticle lifetime leads to increased TLS noise as ${\rm NEP}_{\rm TLS} \propto f_{\rm mod}^{\alpha_{\rm TLS}/2}$ and decreased responsivities through the multiplication of the responsivity $R_x$ (following Eqn. \ref{eqn:NEP_amp}) by $[1 + (2\pi f_{\rm mod} \tau_{\rm qp})^2]^{-1}$, leading to an increase in TLS and amplifier NEP. Note that $\tau_{\rm qp}$ is inversely proportional to the number of quasiparticles in this model. However, for low-quality films or low quasiparticle densities, this quantity may saturate, resulting in a maximum lifetime $\tau_{\rm max}$ \citep{zmuidzinas2012}\done{is it correct to assume then that the analysis presented doesn't include a limitation of $\tau_{qp}$? I would suggest including a statement at the beginning of the paper. TMO: done}. Furthermore, the quasiparticle free-decay time may also limit sensitivities for $f_{\rm mod}$, though $\tau_{\rm qp}$ provides a more stringent constraint for the MKIDs described here. 

With careful accounting for these caveats, the readpower optimization and tone-tracking techniques we describe can provide nearly background-limited sensitivity over a wide range of optical power. This tunability provides an incentive for developing MKIDs for instruments requiring a wide dynamic range, including proposed space telescopes. \repl{}{In the EXCLAIM band, the simpler technique of setting $P_{\rm read}$ to $6\,\rm dB$ below the bifurcation approximates the optimal noise performance on all but the darkest loads.} The widely-ranging background anticipated for the EXCLAIM mission will offer a testbed for these techniques.

\textbf{Acknowledgements} This work was supported by a 5-year NASA Astrophysics Research and Analysis (APRA 17-APRA17-0077) grant and NASA-Goddard Internal Research and Development funds, and TMO acknowledges support from the NASA-Goddard internship program and the UW-Madison graduate program.

\bibliographystyle{abbrvnat}
\bibliography{refs}

\end{document}